\documentclass[aps,prl,superscriptaddress,twocolumn,10pt]{revtex4-1}

\usepackage{graphicx}
\usepackage{amsmath}
\usepackage{dsfont}
\usepackage{xspace}
\usepackage{color}
\usepackage{fixmath}
\usepackage[utf8]{inputenc}

\newcommand{\Rvec}{\mathbf{R}}

\newcommand{\ahatdag}{\hat{a}^\dagger}
\newcommand{\ahat}{\hat{a}}
\newcommand{\nhat}{\hat{n}}

\begin{document}

\title{Microscopy of many-body states in optical lattices}

\author{Christian Gross}%
\affiliation{Max-Planck-Institut f\"{u}r Quantenoptik, 85748 Garching, Germany}
\author{Immanuel Bloch}%
\affiliation{Max-Planck-Institut f\"{u}r Quantenoptik, 85748 Garching, Germany}
\affiliation{Ludwig-Maximilians-Universit\"{a}t, Fakult\"{a}t f\"{u}r Physik, 80799 M\"{u}nchen, Germany}%

\begin{abstract}
  Ultracold atoms in optical lattices have proven to provide an extremely clean and controlled setting to explore quantum many-body phases of matter.
  Now, imaging of atoms in such lattice structures has reached the level of single-atom sensitive detection combined with the highest resolution down to the level of individual lattice sites.
  This has opened up fundamentally new opportunities for the characterization and the control of quantum many-body systems.
  Here we give a brief overview of this field and explore the opportunities offered for future research.
\end{abstract}

\maketitle

\section{Introduction}
\label{sec:introduction}
Over the past years, ultracold atoms in optical lattices have emerged as versatile new system to explore the physics of quantum many-body systems.
On the one hand they can be helpful in gaining a better understanding of known phases of matter and their dynamical behavior, on the other hand they allow one to realize completely novel quantum systems that have not been studied before in nature \cite{Jaksch:2005,Bloch:2008c,Lewenstein:2007}.
Commonly, the approach of exploring quantum many-body systems in such a way is referred to as ``quantum simulations''.
Examples of some of the first strongly-interacting many-body phases that have been realized both in lattices and in the continuum include the quantum phase transition from a Superfluid to a Mott insulator \cite{Fisher:1989,Jaksch:1998,Greiner:2002a}, the achievement of a Tonks-Girardeau gas \cite{Paredes:2004,Kinoshita:2004} and the realization of the BEC-BCS crossover in Fermi gas mixture \cite{RanderiaZwergerZwierlein:2012} using Feshbach resonances \cite{Chin:2010a}.

In almost all of these experiments, detection was limited to time-of-flight imaging or more refined derived techniques that mainly characterized the momentum distribution of the quantum gas \cite{Bloch:2008c}.
However, quantum optics experiments on single or few atoms or ions had shown how powerful the detection and control of individual quantum particles can be.
For several years, researchers in the field have therefore aspired to employ such single particle detection methods for the analysis of ultracold quantum gases.
Only recently it has become possible to realize such imaging techniques, marking a milestone for the characterization and manipulation of ultracold quantum gases \cite{Nelson:2007,Gericke:2008,Bakr:2009,Bakr:2010,Sherson:2010}.
In our discussion, we will focus on one the most successful of these techniques based on high-resolution fluorescence imaging.
Despite being a rather new technique, such quantum gas microscopy has already proven to be an enabling technology for probing and manipulating quantum many-body systems.
For the first time controllable and strongly interacting many-body systems, as realized with ultracold atoms, could be observed on a local scale~\cite{Sherson:2010,Bakr:2010}.
The power of the technique became even more apparent with the advent of local hyperfine state specific addressing in optical lattices~\cite{Weitenberg:2011}.
Together with the local detection this provides a complete toolbox for the manipulation of one- and two-dimensional lattice gases on the scale of a few hundred Nanometers.
Our review will concentrate on the experiments employing ultracold bosonic atoms in optical lattices, showing applications in correlation measurements \cite{Endres:2011,Cheneau:2012,Endres:2013}, quantum magnetism \cite{Fukuhara:2013,Fukuhara:2013a,Simon:2011} and quantum criticial behaviour close to a quantum phase transition \cite{Endres:2012}.

\begin{figure}[b]
\begin{center}
  \includegraphics[width=1\columnwidth]{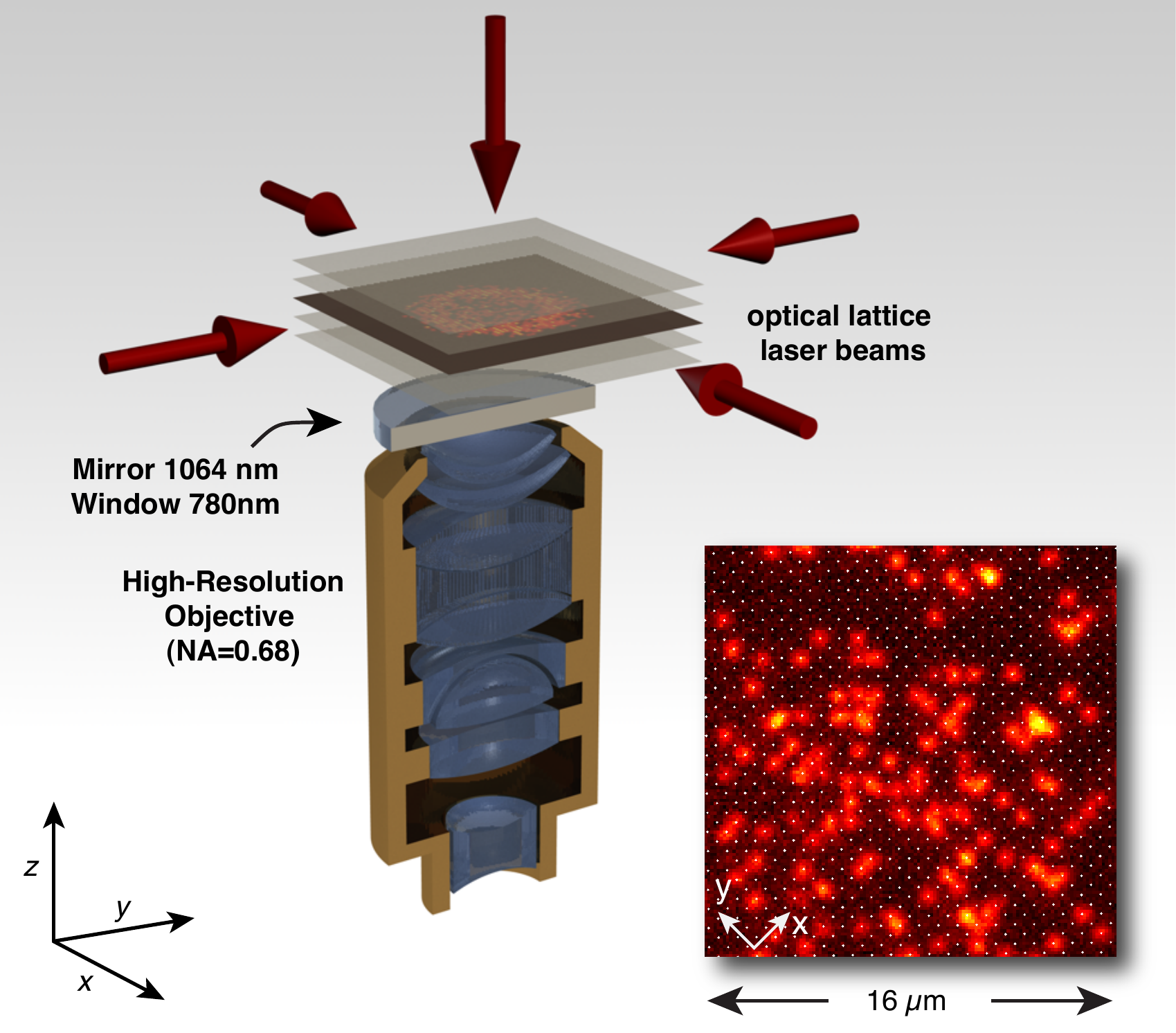}
\end{center}
\caption{\textbf{Schematic setup for high resolution fluorescence imaging of a 2D quantum gas.}
Two-dimensional bosonic quantum gases are prepared in a single 2D plane of an optical standing wave along the $z-$direction, which is created by retro-reflecting a laser beam ($\lambda=1064$\,nm) on the coated vacuum window.
Additional lattice beams along the $x$- and $y$-directions are used to bring the system into the strongly correlated regime of a Mott insulator.
The atoms are detected using fluorescence imaging via a high resolution microscope objective.
Fluorescence of the atoms was induced by illuminating the quantum gas with an optical molasses that simultaneously laser cools the atoms.
The inset shows a section from a fluorescence picture of a dilute thermal cloud (points mark the lattice sites).
Adapted from Sherson et al.~\cite{Sherson:2010}.\label{fig:schematic}}
\end{figure}

\section{Site resolved imaging}
\label{sec:site_resolved_imaging}

One of the standard imaging techniques in ultracold quantum gases -- absorption imaging -- cannot be easily extended to the regime of single atom sensitivity.
This is mainly due to the limited absorption a laser beam experiences when interacting with a single atom.
For typical experimental conditions, the absorption signal is always smaller than the accompanying photon shot noise.
While high resolution images of down to $1\mu$m resolution have been successfully used to record {\em in-situ} absorption images of trapped quantum gases \cite{Gemelke:2009}, they have not reached the single-atom sensitive detection regime.
Fluorescence imaging can however overcome this limited signal-to-noise and therefore provides a viable route for combining high-resolution imaging with single-atom sensitivity.
By using laser induced fluorescence in an optical molasses configuration and by trapping the atoms in a very deep potential, several hundred thousand photons can be scattered from a single atom, of which a few thousand are ultimately detected.
An excellent signal-to-noise in the detection of a single atom can therefore be achieved.

This idea was first pioneered for the case of optical lattices by the group of D. Weiss, who loaded atoms from a magneto-optical trap into a three-dimensional lattice with a lattice constant of 6\,$\mu$m \cite{Nelson:2007}.
However, for typical condensed matter oriented experiments, such large spaced lattices are of limited use, due to their almost vanishing tunnel coupling between neighboring potential wells.
Extending fluorescence imaging to a regime where the resolution can be comparable to typical sub-micron lattice spacings, thus requires large numerical apertures (NA) microscope objectives, as the smallest resolvable distances in classical optics are determined by $\sigma=\lambda/(2 {\rm NA})$.

\begin{figure}[t]
\begin{center}
\includegraphics[width=1\columnwidth]{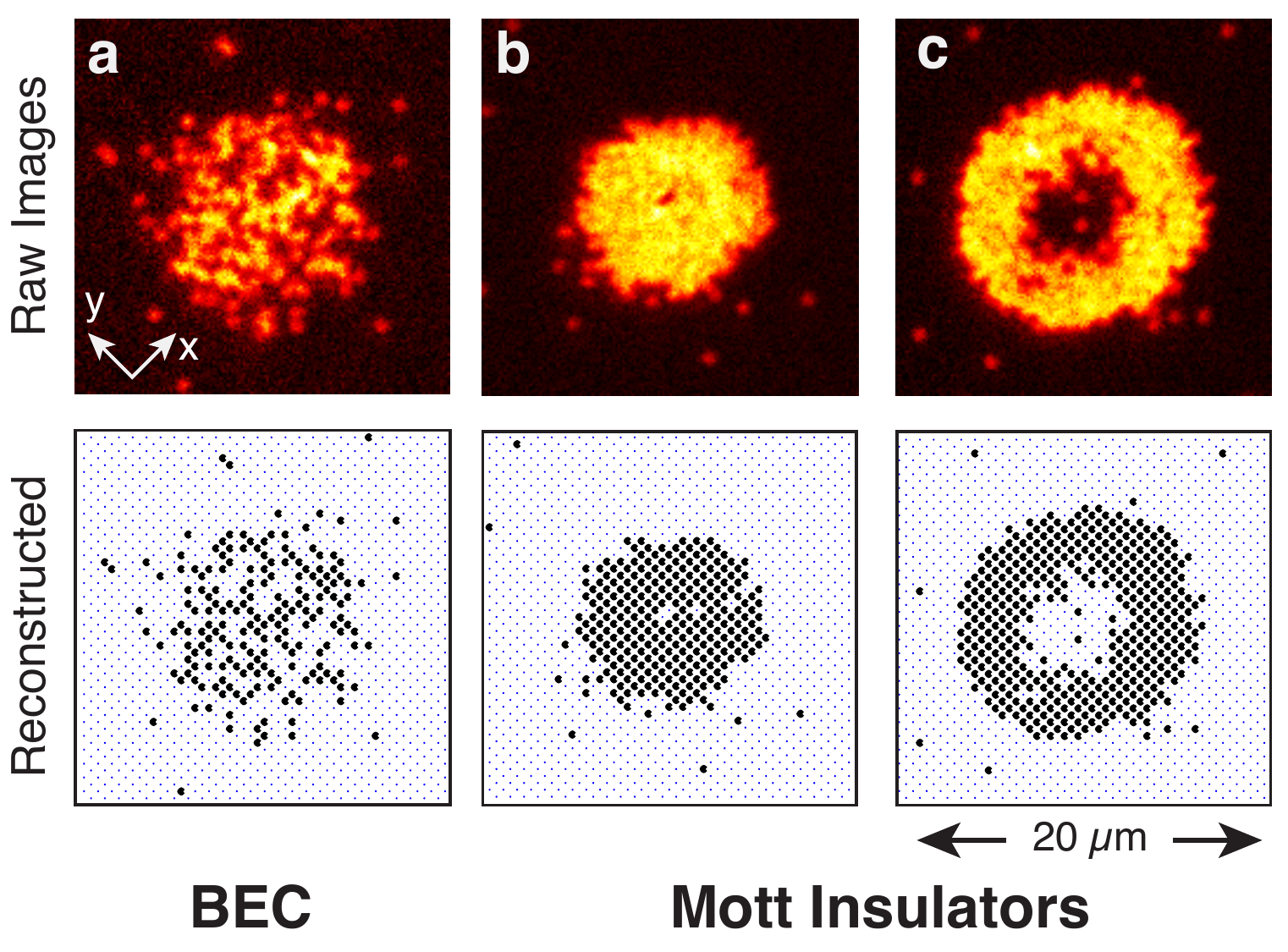}
\end{center}
\caption[]{\textbf{High resolution fluorescence images of a weakly interacting Bose-Einstein condensate and Mott insulators.}
{\bf(a)} Bose-Einstein condensate exhibiting large particle number fluctuations and {\bf(b,c)} wedding cake structure of $n=1$ and $n=2$ Mott insulators.
Using a numerical algorithm, the corresponding atom distribution on the lattice can be reconstructed.
The reconstructed images can be seen in the row below (small points mark lattice sites, large points mark position of a single atom).
Figure adapted from Sherson et al.~\cite{Sherson:2010}.}
\label{fig:bec_mi_series}
\end{figure}

In recent works, Bakr et al. \cite{Bakr:2009,Bakr:2010} and Sherson et al. \cite{Sherson:2010} have demonstrated such high-resolution imaging and applied it to image the transition of a superfluid to a Mott insulator in 2D.
In the experiments, 2D Bose-Einstein condensates were first created in tightly confining potential planes.
Subsequently, the depth of a two-dimensional simple-cubic type lattice was increased, leaving the system either in a superfluid or Mott insulating regime.
The lattice depths were then suddenly increased to very deep values of approximately $300 \mu$K, essentially freezing out the density distribution of the atoms in the lattice.
A near-resonant optical molasses was then used to induce fluorescence of the atoms in the deep lattice and also provide laser cooling, such that atoms remained on lattice sites while fluorescing.
High resolution microscope objectives with numerical apertures of ${\rm NA} \approx 0.7-0.8$ were used to record the fluorescence and image the atomic density distribution on CCD cameras (see Fig.~\ref{fig:schematic}).
A limitation of the detection method is that so called inelastic light-induced collisions occurring during the illumination period only allow one to record the parity of the on-site atom number.
Whenever pairs of atoms are present on a single lattice site, both atoms are rapidly lost within the first few hundred microseconds of illumination, due to a large energy release caused by radiative escape and fine-structure changing collisions \cite{DePue:1999}.

In both experiments, high resolution imaging has allowed one to reconstruct the atom distribution (modulo $2$) on the lattice down to a single-site level.
Results for the case of a Bose-Einstein condensate and Mott insulators of such a particle number reconstruction are displayed in Fig.~\ref{fig:bec_mi_series}.
The fidelity of the imaging process is currently limited to approximately $99\%$ by atom loss during the illumination due to background gas collisions.

\section{Thermometry at the limit of individual thermal excitations}
\label{sec:mi_thermodynamics}
Within a tight-binding approximation and for interaction energies that do not exceed the vibrational level splitting on a single lattice site, the behaviour of interacting bosons on a lattice can be described by the Bose-Hubbard Hamiltonian \cite{Fisher:1989,Jaksch:1998,Greiner:2002a}:

\begin{equation}
  \hat H = - J \sum_{\langle \Rvec,\Rvec' \rangle} \ahatdag_{\Rvec'} \ahat_\Rvec + \frac{1}{2} U \sum_\Rvec \nhat_\Rvec (\nhat_\Rvec-1),
\end{equation}

where $\ahatdag_\Rvec (\ahat_\Rvec)$ denote the bosonic creation (annihilation) operators on site $\Rvec$, $\nhat_\Rvec = \ahatdag_\Rvec \ahat_\Rvec$, $J$ characterizes the tunnel coupling between neighbouring lattice sites and $U$ is the on-site interaction energy of two atoms on a given lattice site. For the following discussion, we will assume repulsive interactions $U>0$.

Deep in the Mott-insulating regime the strongly interacting bosonic quantum gas becomes essentially classical. In this so called \textit{atomic-limit of the Bose-Hubbard model} the individual wells are disconnected, that is, the tunneling is $J=0$ and the ratio of interaction to tunneling $U/J$ diverges.
Hence, the grand canonical partition function of the trapped quantum gas $Z^{(0)}$ can be written as a product of on-site partition functions $Z^{(0)} = \prod_\Rvec Z_\Rvec^{(0)}$, where the on-site partition function is given by
\begin{equation}
  Z^{(0)}= \sum_n e^{-\beta(E_n - \mu(\Rvec)n)}\,.
\end{equation}
The local chemical potential at lattice site $\Rvec$ is denoted by $\mu(\Rvec)$ and the eigenenergy of $n$ atoms on this lattice site is given by the standard single-band Bose-Hubbard interaction term $E_n = 1/2 U n (n-1)$.
In particular we can use the above to calculate the on-site probability of finding $n$ atoms per lattice site as
\begin{equation}
  P_\Rvec(n) = \frac{e^{-\beta(E_n - \mu(\Rvec)n)}}{Z^{(0)}}.
\end{equation}
Thus, the thermodynamics is determined only by the ratio of $U/(k_B T)$ and the local chemical potential.
In this limit the problem becomes analytically tractable and simple to analyze.

As a simple application of our result, let us calculate the density profile and its fluctuations for a two-dimensional radially symmetric trapping potential.
All sites with the same distance $r$ from the trapping center exhibit the same chemical potential $\mu(r)$.
The average density at this radial distance is thus given by $\bar n = \frac{1}{Z(r)} \sum_n n e^{-\beta(E_n - \mu(r))}$.
In order to evaluate this, we would need to sum over all possible occupation states in our on-site partition function.
In practice, we may truncate our sum around occupation numbers $n_{\mathrm{max}} = \mathrm{ceil}(\mu/U)+1$ for temperatures $k_B T \ll U$, as thermal fluctuations become exponentially suppressed in this regime.
This corresponds to the so called {\em particle-hole} approximation.
In this regime individual thermal excitations, corresponding either to a missing or and extra atom on a single site are directly detected by the local parity sensitive imaging.

\begin{figure}[t]
\begin{center}
    \includegraphics[width=1\columnwidth]{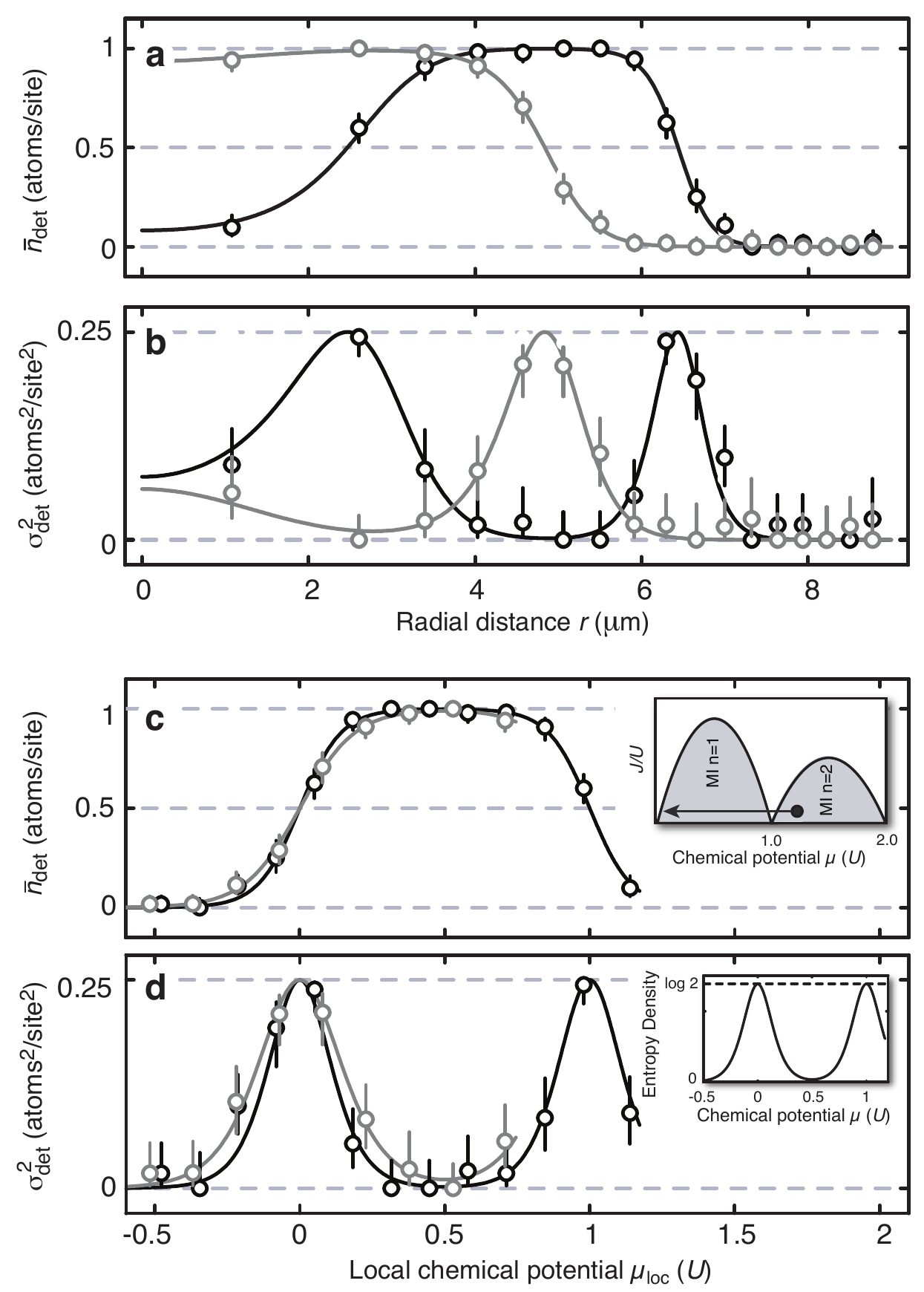}
\end{center}
\caption{{{\bf Radial atom density and variance profiles.} Radial profiles were obtained from the reconstructed images by azimuthal averaging.
{\bf a}, {\bf b}, grey and black points correspond to the $n=1$ and $n=2$ MI images of Fig.\,\ref{fig:bec_mi_series}d,e.
For the two curves, the fits yielded temperatures $T=0.090(5)U/k_B$ and $T=0.074(5)U/k_B$, chemical potentials $\mu=0.73(3)U$ and $\mu=1.17(1)U$, and radii $r_0=5.7(1)\,\mu$m and $r_0=5.95(4)\,\mu$m respectively.
From the fitted values of $T$, $\mu$ and $r_0$, we determined the atom numbers of the system to $N=300(20)$ and $N=610(20)$.
 {\bf c}, {\bf d}, The same data plotted versus the local chemical potential using local-density approximation.
The inset of {\bf c} is a Bose-Hubbard phase diagram ($T=0$) showing the transition between the characteristic MI lobes and the superfluid region.
The line starting at the maximum chemical potential $\mu$ shows the part of the phase diagram existing simultaneously at different radii in the trap due to the external harmonic confinement.
The inset of {\bf d} is the entropy density calculated for the displayed $n=2$ MI.
From Sherson et al.~\cite{Sherson:2010}}\label{fig:Mott_profiles}}
\end{figure}

Taking into the light assisted collisions into account we find for the detected average density:
\begin{equation}
  \bar n_{det}  = \frac{1}{Z(r)} \sum_n \mod_{\!\!2} (n) e^{-\beta(E_n - \mu(r))}.
\end{equation}
The parity projection during the imaging process assures that the experimentally detected atoms number per site is either $0$ or $1$. Thus, the second moment of the measured onsite atom number distribution is equal to its mean  $\overline{n_{det}^2} = \bar n_{det}$. 
Within the particle-hole approximation the physical atom number per site can only fluctuate by $\pm 1$ around its average value, such that its variance $\sigma^2$ can be measured despite parity projection $\sigma^2 = \bar n_{det} - \bar n^2_{det}$.
Both, the average density profile and its fluctuations are functions of three parameters $\mu/U$, $k_B T/U$ and the trapping frequency $\omega$ of the overall harmonic confinement.
While the trap frequency can be independently measured, the chemical potential and temperature of the quantum gas can be extracted via a fit to azimuthally averaged radial density profiles of single images of the quantum gas.
This is shown in Fig.~\ref{fig:Mott_profiles} for the two images of an $n=1$ and $n=2$ Mott insulator in the core of the gas.

In the atomic limit, these fit-functions thus allow an efficient determination of temperature and chemical potential of the quantum gas.
The radial density and fluctuation profiles can be converted to density and fluctuation profiles vs chemical potential by using again the local density approximation $\mu_{loc}(r) = \mu - 1/2 m \omega^2 r^2$.
We see that the data for the two distinct measurements of the $n=1$ and $n=2$ Mott insulators fall on top of each other when plotting in this way, underlining the fact that radial profiles correspond to cuts through the phase diagram (see inset in Fig.~\ref{fig:Mott_profiles}c) of the Bose-Hubbard model.
Residual small differences between the two curves can be attributed to the slightly different temperatures of the atom clouds.
Interestingly, as in the case of fermionic atoms, one notes that fluctuations are concentrated to the border of the Mott insulating regions, where the system is superfluid for $T=0$.
In the local density approximation picture, the energy gap is minimal here and, thus, it is thermodynamically easiest to introduce fluctuations in this spatial region.
Note that this thermometry method can be extended to work also in the vicinity of the superfluid to Mott insulator transition.
In this case the local quantum fluctuations can be suppressed prior to the imaging by a properly timed lattice ramp that ends deep in the Mott insulator regime~\cite{Endres:2012}.

\section{Single-site resolved addressing of individual atoms}
\label{sec:single_atom_addressing}

Being able to spatially resolve single lattice sites also allows to manipulate atoms with single-site resolution.
A laser beam can be sent in reverse through the high-resolution objective and, hence, is focused onto the atoms.
Thereby the high-resolution objective is used twice -- for imaging and for local addressing.
In typical cases, the resulting spot size of the laser beam will still be on the order of a lattice spacing and for most applications too large in order to reliably address atoms on single lattice sites.
One possibility to increase the spatial resolution is to make use of a resonance imaging technique: the focused laser is tuned to such a wavelength that it creates a differential energy shift between two internal hyperfine ground states of an atom.
Then global microwave radiation will be resonant only at the position of the focused beam and thus, can be used to control the spin state of the atom \cite{Weiss:2004,Weitenberg:2011}.
The spatial resolution for the addressing of single atoms can thereby be increased up to a limit given by (often magnetic field driven) fluctuations of the energy splitting between the two hyperfine states.
For typical parameters this corresponds to an increase by almost an order of magnitude down to $\simeq 50$\,nm, well below the optical diffraction limit.

\begin{figure}[t]
\begin{center}
    \includegraphics[width=1\columnwidth]{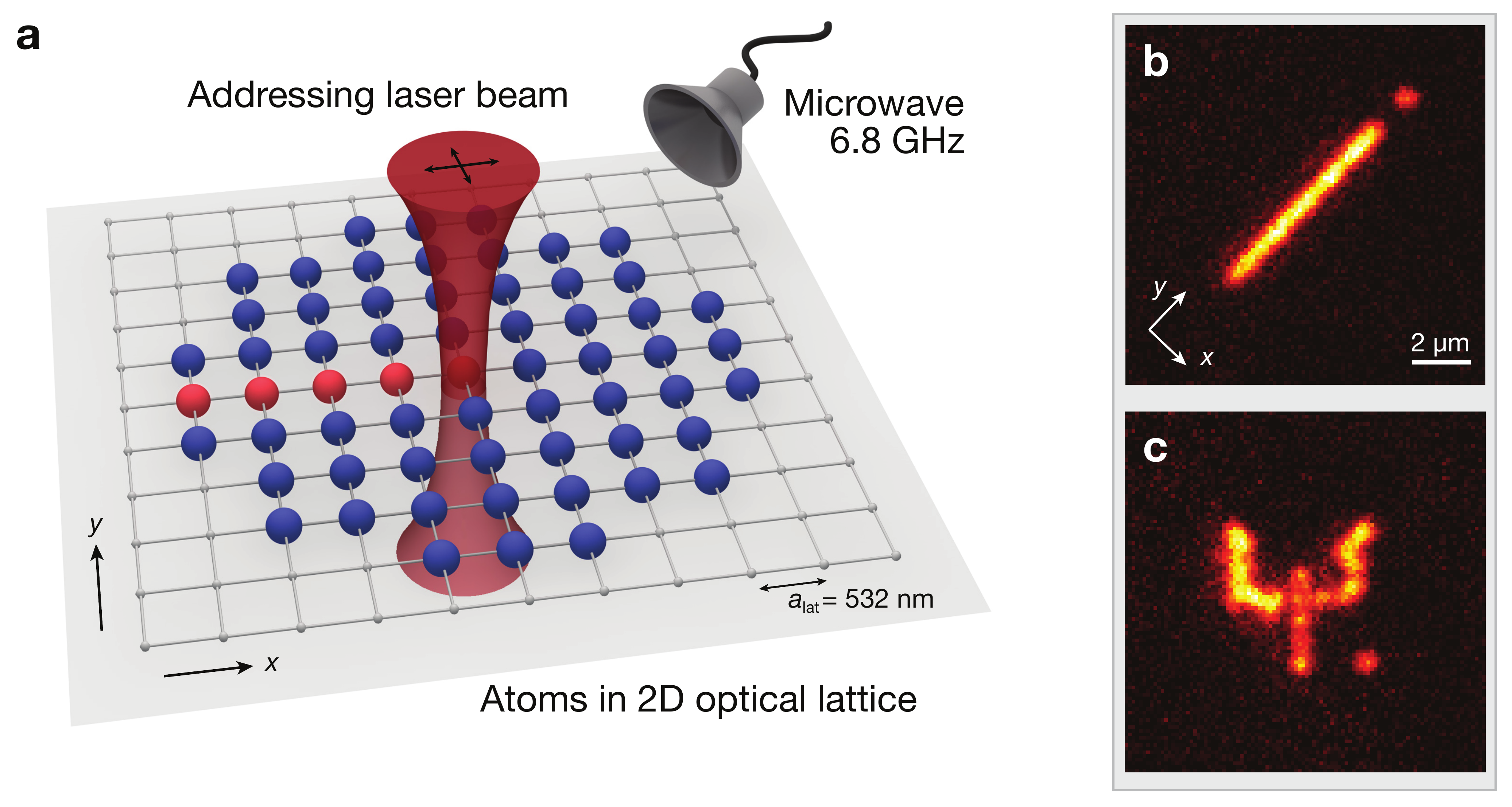}
\end{center}
\caption{{{\bf High-resolution addressing of single atoms.} {\bf (a),} Atoms in a Mott insulator
with unity filling arranged on a square lattice with period $a_{\text{lat}}=532$\,nm were addressed using an off-resonant laser beam. The beam was focused onto individual lattice sites by a high-aperture microscope objective (not shown) and could be moved in the $xy$ plane with an accuracy of better than $0.1\,a_{\text{ lat}}$.
 {\bf (b,c)}} Fluorescence images of spin-flipped atoms following the addressing procedure.
From Weitenberg et al.~\cite{Weitenberg:2011}.\label{fig:addressing}}
\end{figure}

In the experiment, such addressing was demonstrated in a 2D Mott insulator with unity occupation per lattice site \cite{Weitenberg:2011}.
In order to prepare an arbitrary pattern of spins in the array, the laser beam was moved to a specific site and a Landau-Zener microwave sweep was applied in order to flip the spin of the atom located at the lattice site.
The laser beam was then moved to the next lattice site and the procedure was repeated.
In order to detect the resulting spin pattern, unflipped atoms were removed by applying a resonant laser beam that rapidly expelled these atoms from the trap \cite{Weitenberg:2011}.
The remaining spin-flipped atoms were then detected using standard high-resolution fluorescence imaging, as described above.
The resulting atomic patterns can be seen in Fig.~\ref{fig:addressing}, showing that almost arbitrary atomic orderings can be produced in this way.
The described scheme can be enhanced to allow for simultaneous addressing of multiple lattice sites using an intensity shaped laser beam instead of a focused Gaussian beam.
Such a beam can be prepared in the lab using spatial light modulators~\cite{Fukuhara:2013}.

\begin{figure*}[t]
\begin{center}
    \includegraphics[width=0.7\textwidth]{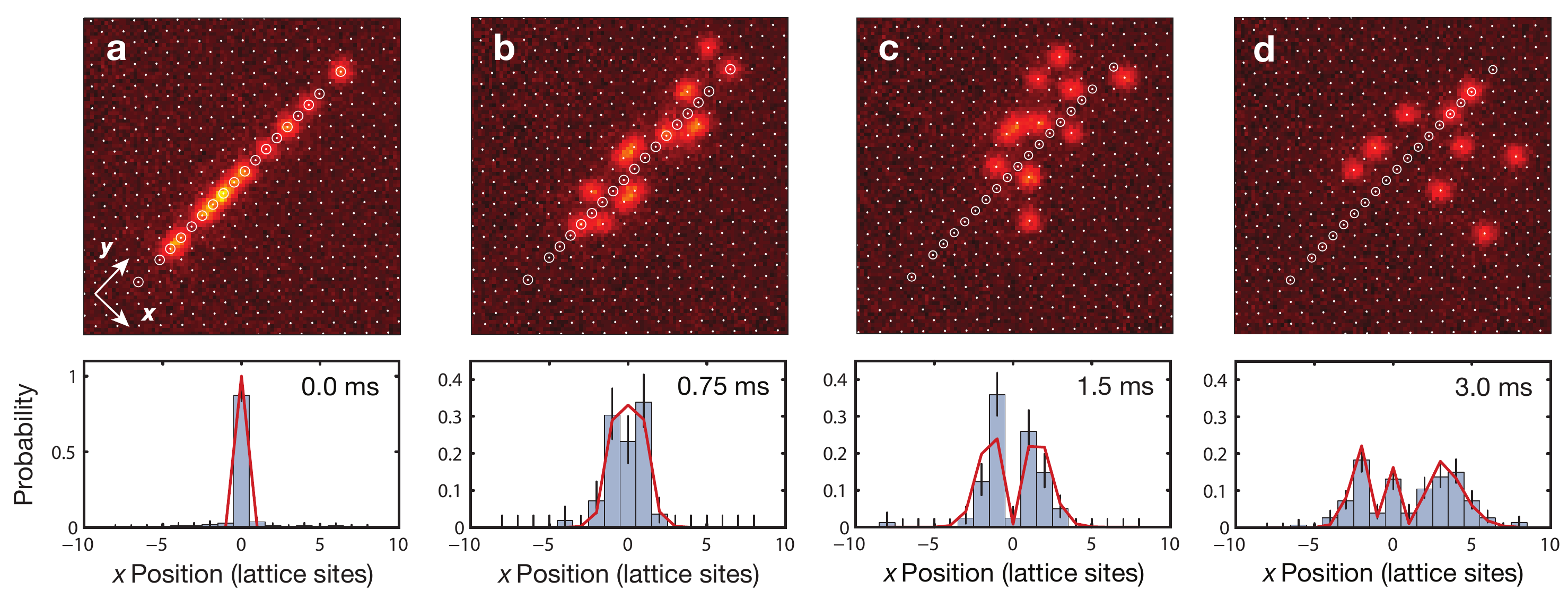}
\end{center}
\caption{{{\bf Tunneling of single particles on a lattice.} {\bf (a),} Atoms were prepared in a single line along the $y$ direction before the lattice along the $x$ axis was lowered, allowing the atoms to tunnel in this direction {\bf (b-d)}.
The top row shows snapshots of the atomic distribution after different hold times.
White circles indicate the lattice sites at which the atoms were prepared (not all sites initially contained an atom).
The bottom row shows the respective position distribution obtained from an average over $10-20$ of such pictures, the error bars give the $1\sigma$ statistical uncertainty.
The red curce corresponds to the prediction by theory.
From Weitenberg et al.~\cite{Weitenberg:2011}.}\label{fig:tunneling}}
\end{figure*}

In order to demonstrate that the addressing does not affect the motional state of the atoms on the lattice site, the tunneling of particles was investigated after an addressing sequence.
Using the addressing sequence described above, a line of atoms in $y$ direction was prepared from a Mott insulator in a deep lattice.
Thereafter, the lattice depth along the $x$-direction was lowered in order to initiate tunneling of the particles along this direction.
After a variable evolution time, the position of the atoms was measured (see Fig.~\ref{fig:tunneling}).
By repeating the experiment several times, the probability of finding the atom at a certain lattice site for a specific evolution time could be determined and compared to the probability distribution predicted by the Schr\"odinger equation for the quantum evolution of a single particle tunneling on a lattice.
Excellent agreement was found between the experimental data and the theoretical prediction, indicating that most atoms indeed were still in the lower energy band of the lattice despite the addressing.
Atoms in higher energy bands typically exhibit an order of magnitude larger tunnel coupling, allowing them to travel much further given the same evolution time.
However, in the experiment a negligible fraction of atoms was detected at such positions in the experiment.

High resolution imaging and addressing can be very useful for preparing almost arbitrary initial configurations of the many-body system that can e.g. be used to investigate a specific non-equilibrium evolution.
It can also be highly beneficial for quantum information applications, where e.g. in the case of a one-way quantum computer \cite{Raussendorf:2001}, it is essential to measure the spin state of an atom at a specified lattice site.

\section{Quantum gas microscopy -- an enabling technology}
\label{sec:quantum_gas_microscopy_an_enabling_technology}

Combining the techniques described above, quantum gas microscopy has proven to be an enabling technology for probing and controlling quantum many-body systems.
The imaging method allows for the measurement of local counting statistics of the atomic parity for strongly correlated many-body states.
For example, correlation functions -- not necessary restricted to two-point correlators -- can be extracted from the data~\cite{Endres:2013,Rath:2013}.
Based on those, the \textit{quantum melting} of one- and two-dimensional Mott insulators through a proliferation of correlated particle-hole pairs has been directly observed.
Furthermore, non-local multi-point correlators have been extracted to analyze the emerging string order, a hidden order parameter for Mott-insulating states at zero temperature~\cite{Endres:2011}.
In the context of topologically ordered phases of matter, non-local order parameters play a crucial role to characterize the complex entanglement order present in these states \cite{Wen:2004,DallaTorre:2006,Anfuso:2007a}.
So far, it was believed that non-local order is merely a theoretical concept, not accessible to experiments.
Quantum gases microscopy now makes probing such highly non-trivial order a reality for experiments.
The local parity sensitive detection is also ideally suited to study low lying excitations of the strongly correlated system close to the Mott-insulating phase.
Here, the excitations in the system can be converted into particle-hole excitations by a sweep of the lattice depth such that the final state is deep in the Mott-insulating regime.
In this regime, the detection scheme is sensitive to single quasiparticles (i.e. these particle-hole excitations) such that bolometric measurements with highest sensitivity are possible.
Such measurements enabled the detection of a mode softening around the particle-hole symmetric critical point in two-dimensions, which could be attributed to a Higgs-like excitation on the superfluid side of the transition~\cite{Endres:2012}.

\begin{figure*}[t]
  \begin{center}
    \includegraphics[width=0.6\textwidth]{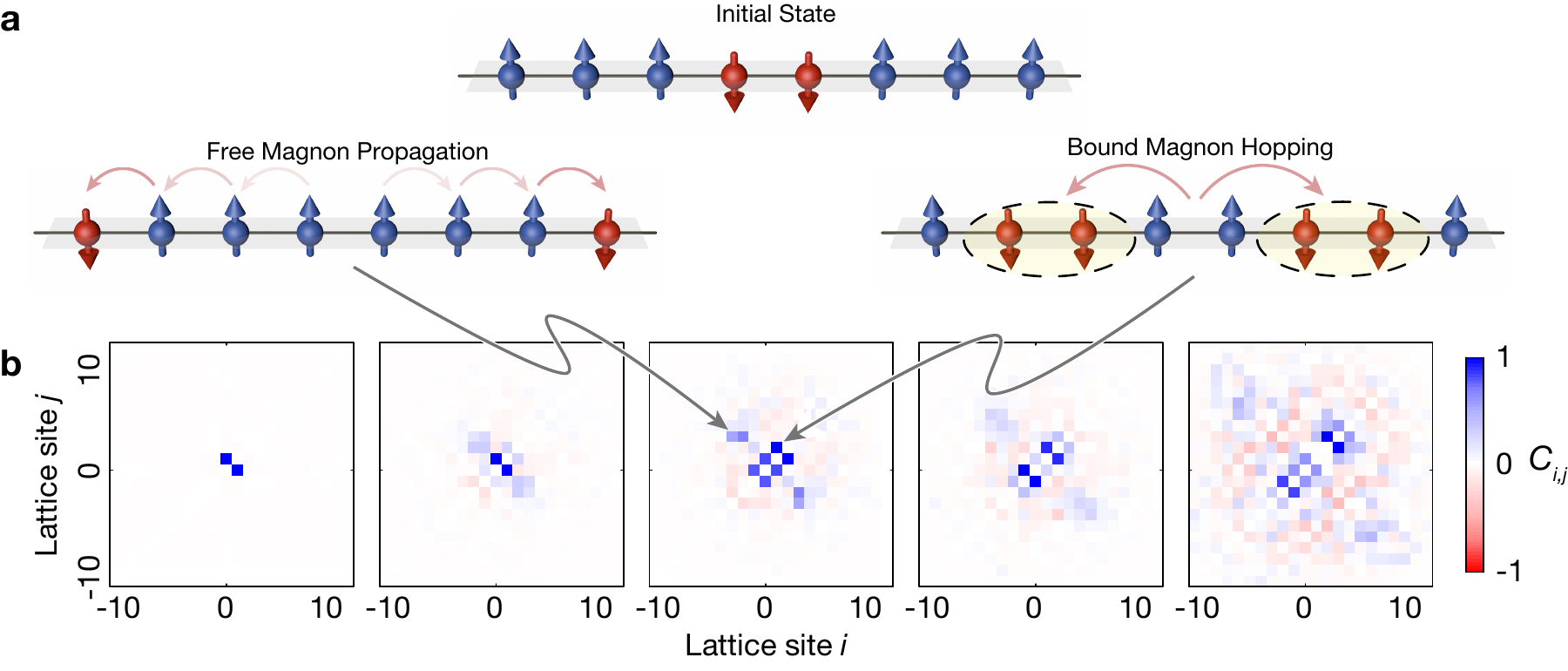}
  \end{center}
  \caption{\textbf{Magnon dynamics.} \textbf{(a), } Initially two adjacent spins are flipped in an otherwise fully polarized Heisenberg chain.
This state has roughly equal overlap with \textit{free} magnon and \textit{bound} magnon states.
During the subsequent evolution the free and bound magnon contributions develop distinct features that are revealed in correlation measurements shown in \textbf{(b)}.
The plots show the correlation (color scale) between two sites for $0\,$ms, $40\,$ms, $60\,$ms, $80\,$ms and $120\,$ms from left to right.
The bound magnons tend to stick together leading to a high nearest neighbor correlation signal on the ``lower left to upper right'' diagonal for all times.
On the contrary, the free magnons show anti-bunching behavior, maximizing their distance.
This shows up as a signal in the ``upper left to lower right'' diagonal.
Adapted from Fukuhara et al.~\cite{Fukuhara:2013a}.\label{fig:magnons}}
\end{figure*}

Next to equilibrium physics, also dynamical properties of strongly correlated systems in optical lattices can be studied.
This is especially remarkable, since it allows for real-time tracking of the dynamics in the system.
A controlled quench of the lattice height of a one-dimensional lattice gas into the Mott-insulating state excites the energetically low-lying particle-hole excitations homogeneously within the system.
These excitations manifest themselves in characteristic correlations based on entangled quasiparticles that spread out across the system with a fixed velocity.
The light-cone like spreading of correlations, first predicted by Lieb-Robinson~\cite{Lieb:1972}, could thereby be revealed for the first time experimentally~\cite{Cheneau:2012}.

The hyperfine state selective microscopic detection and manipulation technique is ideally suited to study bosonic quantum magnetism in optical lattices.
Heisenberg type magnetic couplings can be implemented by using two internal hyperfine degrees of freedom, on which a pseudo-spin $1/2$ is defined.
For such systems, the anisotropy in the spin couplings can in principle be controlled either by Feshbach resonances or by state dependent hoppings~\cite{Kuklov:2003,Duan:2003,Altman:2003}.
In the case of Rubidium and spin independent lattices the resulting Heisenberg Hamiltonian is -- up to a few percent -- symmetric in the spin coupling.
It takes the simple form
\begin{equation}
  \hat{H} = -J_{\mathrm{ex}} \sum_i \mathbf{\hat{S}}_i \cdot \mathbf{\hat{S}}_i\, ,
\end{equation}
where $J_{\mathrm{ex}}$ is the superexchange coupling $J_{\mathrm{ex}} = 4 J^2/U$.
This superexchange coupling describes spin exchange processes obtained from second order perturbation terms in the Mott insulating phase at unity filling.
This is commonly referred to as the \textit{strong-coupling limit}~\cite{Kuklov:2003,Duan:2003}.
In first experiments, direct observation of superexchange couplings~\cite{Trotzky:2008a} and detection of singlet-triplet spin correlations in double wells~\cite{Trotzky:2010,Greif:2013} as well as more complex plaquette resonating valence bond states~\cite{Paredes:2008,Nascimbene:2012} were observed.
Now, using quantum gas microscopes, the detection and control possibilities for quantum magnetism have also been dramatically enhanced.

When characterizing the spin-spin exchange couplings, one finds that already in one-dimensional systems the corresponding energy scale is small $J_\mathrm{ex} = h\times \mathcal{O}(10\,\mathrm{Hz})$.
In higher dimensions an even smaller ratio $J/U$ is required to reach the Mott insulating phase resulting in an even more reduced exchange coupling~\cite{Kashurnikov:1996,Kuhner:2000,Capogrosso-Sansone:2007,Capogrosso-Sansone:2008}.
These tiny energy scales pose a major open challenge to observe characteristic magnetic quantum correlations in thermal equilibrium, as temperature is typically larger than such exchange couplings.
However, in a spin-polarized Mott insulator, entropy is not distributed uniformly throughout the system, but is rather confined to narrow regions at the boundary of the system (see section \ref{sec:mi_thermodynamics}).
The core of such a fully polarized Mott insulator can therefore be regarded to be at almost zero temperature, forming an ideal initial state for the observation of coherent quantum magnetic phenomena.
Especially, in combination with high fidelity local addressing, this allows for the deterministic preparation of precisely controlled initial spin distributions, whose ensuing quantum evolution can be readily tracked.
Using such a technique, the coherent dynamics of a single magnetic quasiparticle, a magnon, could be observed in Heisenberg spin chains~\cite{Fukuhara:2013}.
These measurements were carried out in the subspace of a single spin impurity, such that the next neighbor spin-interaction term $\propto S^z_i \, S^z_j$  does not play any role for the dynamical evolution.

Given this ultimate control over the initial local magnetization, complexity can be added step wise to the problem.
The simplest setting in which the magnetic interaction, i.e. the $\hat{S}^z_i \hat{S}^z_{i+1}$ coupling, becomes important is the case of two spin impurities on the Heisenberg chain.
This scenario can be readily studied by flipping two adjacent spins in the initially fully polarized chain.
Such a state has overlap both with free magnon as well as bound magnon states and one therefore expects to observe both propagation phenomena in the subsequent dynamical evolution of the initial state.
The emergence of the low energy bound states in the excitation spectrum is probably the most striking  microscopic effect of ferromagnetism~\cite{Bethe:1931,Wortis:1963} and in fact can be seen as the most elementary magnetic soliton.
These bound states have recently been directly observed and characterized by site resolved correlation measurements~\cite{Fukuhara:2013a}.
Their signature, a high probability of finding the two impurity spins on adjacent sites even after a long time, can be seen in figure~\ref{fig:magnons}.

Studies of quantum magnetism is not limited to the symmetric Heisenberg scenario described above.
Quantum Ising models, that is, models with classical Ising coupling in addition to transverse single spin couplings by external fields, can be achieved by a different, spatial encoding of the pseudo spin that is defined on the bonds of a tilted lattice, i.e. in between two sites~\cite{Sachdev:2002}.
The latter technique maps an empty site next to a doubly occupied site to one of the two spin states and two singly occupied states next to each other to the other one.
Thus, parity resolved local detection is an ideal tool to study Ising spin chains using this mapping as realized in M. Greiner's group~\cite{Simon:2011}.

\section{Outlook}
\label{sec:outlook}

The novel techniques to image and control individual atoms that we have outlined above mark a milestone in the experimental control over quantum many-body systems.
In fact we believe that we have only scratched the surface of the wide range of applications that will emerge in the future.
The possibility to reveal hidden order parameters of topological phases of matter in higher dimensions~\cite{Wen:2004,Rath:2013}, the ability to measure the full counting statistics in a many-body setting or the possibility to directly measure entanglement entropies~\cite{Daley:2012,Pichler:2013} will open new avenues for our understanding of correlated quantum phases of matter.
Currently, work is progressing in several groups to realize such imaging and control also for fermionic quantum gases.
There, it could be directly used to characterize long-ranged magnetic correlations~\cite{Lee:2006} in the fermionic Hubbard model and to excite individual magnetic quasiparticles in the system and observe their propagation.
Single atom imaging has now also found useful applications in the field of Rydberg gases, where the imaging of crystalline structures of an ensemble of Rydberg atoms has only been possible due to these new detection capabilities~\cite{Schauss:2012,Schauss:2014}.
Extending the control towards individual Rydberg atoms might enable one to realize novel multi-particle interactions in a Rydberg quantum simulator that could be e.g. used to implement exotic spin models such as those underlying Kitaev's toric code \cite{Weimer:2010}, spin-ice models~\cite{Glaetzle:2014a} or even non-Abelian lattice gauge theories~\cite{Tagliacozzo:2013}.
High-resolution imaging will certainly also find other applications, e.g. in realizing the confinement and providing the resolution needed to probe edge state physics in the recently realized topologically band structures in optical lattices \cite{Atala:2013,Aidelsburger:2013,Miyake:2013,Atala:2014,Jotzu:2014,Aidelsburger:2014}.
Next to imaging the local occupation in such states, it could also be employed to directly probe the particle currents along individual bonds in the lattice \cite{Trotzky:2012,Kessler:2014,Atala:2014} thereby providing fundamentally complementary information to measurements in the occupation basis.
While technically challenging to implement, we believe that the imaging and control techniques outlined here offer so many advantages and potential that they will in fact become standard techniques on any cold atom or molecule experiment in the future.

\bibliography{manuscript}
\end{document}